\documentclass[prd,showpacs,showkeys]{revtex4} 
\usepackage{amsmath,amssymb,bm}
\usepackage{epsfig}
\usepackage{graphicx}
\usepackage{amsfonts}
\usepackage{epstopdf}
\usepackage[usenames,dvipsnames]{color} 
\usepackage{float}
\usepackage{booktabs}
\usepackage{multirow}
\newcommand{\nn}{\nonumber}
\newcommand{\beqa}{\begin{eqnarray}}
\newcommand{\eeqa}{\end{eqnarray}}

\def\ii{\'i}
\begin{document}

\title{
Theoretical estimates of the width of light-meson states in the SO(4)
(2+1)-flavor limit.}

\author{Tochtli Y\'epez-Mart\ii nez}
\email{yepez@fisica.unlp.edu.ar}
\author{Osvaldo Civitarese}
\email{osvaldo.civitarese@fisica.unlp.edu.ar}
\affiliation{Departamento de F\ii sica, Universidad Nacional de La Plata, C.C.67 (1900), La Plata, Argentina}  
\author{Peter Otto Hess}
\email{hess@nucleares.unam.mx}
\affiliation{Instituto de Ciencias Nucleares, Universidad Nacional Aut\'onoma de M\'exico, 
Ciudad Universitaria, Circuito Exterior S/N, A.P. 70-543, 04510 M\'exico D.F. M\'exico}  

\date{\today}

\begin{abstract}
The low-energy sector of the mesonic spectrum exhibits some features which may be understood in terms of the SO(4) symmetry 
contained in
the QCD-Hamiltonian written in the Coulomb Gauge. In our previous work we have shown that this is indeed the case when the 
Instantaneous Color-Charge Interaction (ICCI) is treated by means of non-perturbative many-body linearization techniques.
Continuing along this line of description in this work we calculate the width of meson states belonging to the low portion of 
the spectrum (E $<$ 1 GeV ). In spite of the rather simple structure of the Hamiltonian used to calculate the spectra of pseudoscalar and 
vector mesons, the results for the width of these states follow the pattern of the data.   
\end{abstract}
\maketitle

\section{Introduction} 
\label{Sec: introduction}

In \cite{Yepez2016}, a QCD motivated Hamiltonian for light quarks was
introduced. It was shown that a description of low-lying physical meson
states can be given in terms of the eigenstates of the Casimir operator of the SO(4) group, since a sector
of the QCD Hamiltonian possesses such a symmetry. In the same work \cite{Yepez2016} it was shown that the pion-like meson 
state is an eigenstate of the Casimir operator, of the singlet SO(4) representation, with zero energy.

In \cite{Yepez2017} the chiral and flavor symmetries, present in the above mentioned SO(4) limit of the QCD-Hamiltonian 
discussed in \cite{Yepez2016}, were
broken and a (2+1)-flavor description for light and strange quarks was used. 
The effective Hamiltonian was diagonalized in a basis of
quark-antiquark pairs by applying the {\it Random Phase Approximation} (RPA) method.
The eigenvalues shown a correspondence with the spectrum of light
meson states (E $<$ 1 GeV)
but since the SO(4) symmetry was only partially broken the
eigenstates could only be interpreted by their dominant flavor content
and  their energies. The effects of the ground state correlations, accounted for 
in the RPA scheme, become evident in the collectivity of
certain states like the pion-like states and the $\eta$- and
$\eta'$-like states \cite{Amor2017,llanes}. The quark-content of each of the RPA solutions was analysed and the states
where labelled by their likelihood with physical states.

In \cite{Amor2017} the SO(4) formalism was extended to the level of the Coulomb-Gauge-QCD-Hamitonian \cite{Lee1}.
The Instantaneous-Color-Coulomb Interaction (ICCI) was replaced by  
a confining Coulomb-plus-linear potential. The resulting eigenvalue-problem
was solved by applying many-body methods, like the RPA method. 
Since the calculations were performed in a larger basis, as compared with \cite{Yepez2016,Yepez2017}, the structure of pseudoscalar, vector
and scalar meson-like states was obtained and compared to the available data.
It was found that the RPA spectrum reproduced several
characteristics of the experimental meson spectrum up to 1 GeV.
The analysis of \cite{Amor2017} could be extended to include the calculation of the width of the states, 
which could provide relevant information about the nature of the observed states, and this 
is the main purpose of the present work.
These observables are indeed very relevant to the understanding of the structure of low-energy meson-states, since
large values of the widths have been reported for the case of scalar and vector mesons 
while very small values of the order of few eV to
$1.5~\mbox{MeV}$ are assigned to pseudoscalar mesons \cite{PDG2016}.
In the present work we calculate
the width of meson-like states in the basis of effective degrees of freedom contained in the SO(4)
model of \cite{Yepez2016,Yepez2017}. Although the model is rather simple it has the advantage of a relatively small number of 
effective degrees of freedom, a feature which facilitates the identification of physical states by their quark contents.    

The  paper is organized as follows. In Section \ref{Sec2} we briefly describe the essentials of the QCD-SO(4) model, introduce 
its RPA treatment and the formalism used to calculate the width of the states.
In Section \ref{Sec3} we present and discuss the results of the calculations.  
Finally, we summarize our conclusions in Section \ref{Sec4}.

\section{Formalism}\label{Sec2}
\subsection{SO(4) model and its RPA solutions.}\label{SO(4) solutions}

A general Hamiltonian for quarks and antiquarks has one-body terms
$H_{11}+H_{20}+H_{02}$ and  two-body terms
$H_{22}+H_{31}+H_{13}+H_{40}+H_{04}$, where the first and second
sub-indexes indicate the number of creation and annihilation operators
appearing in each term, respectively.
In \cite{Yepez2017}, we have implemented the RPA method and found 
meson-like solutions for a Hamitonian of the form,
\beqa\label{HRPA}
H_{RPA}=H_{11}+H_{22}+H_{40}+H_{04}~,
\eeqa
where each term of the Hamiltonian was expressed in terms of the SO(4)-group 
generators obtained for a system of particles and holes described in
\cite{Yepez2016}. 
The Hamiltonian based on the SO(4) generators reads: 

\beqa\label{H-RPA-flavor}
H_{RPA}[SO(4)]&=&
\left(\epsilon_f C^\dag_{2 m,f}C_{2 m,f}- \epsilon_{f'} C^\dag_{1 m,f'}C_{1 m,f'}  \right)    
-a_7\hat V_0  +a_2 \hat V_0^2      +a_3 \hat J^2_0     +a_6\hat V_0 \hat J_0
+\frac{a_1}{2}  \left(\hat J_+ \hat J_- + \hat V_+ \hat V_- 
\right) \nn\\
&+& 
\frac{a_5}{2} \left(\hat J_+ \hat V_- + \hat V_- \hat J_+ +h.c.\right) 
+b\left( (\hat J_++\hat V_+) (\hat J_++\hat V_+) +h.c.\right)~.
%+b_1\left(  \hat J_+^2 + \hat V_+^2 +  \hat J_-^2 + \hat V_-^2\right) 
%+b_2\left(  \hat J_+ \hat V_+ +\hat V_+ \hat J_+  + \hat J_- \hat V_- +\hat V_- \hat J_-\right) ~.
\eeqa 

Four sets of parameters were used for the
coefficients $(a_i,b)$ of the Hamiltonian of Eq.(\ref{H-RPA-flavor}) leading to four different
scenarios for the low energy meson-like spectrum \cite{Yepez2016}.
The sets of parameters and the corresponding solutions were denoted by
{\it Set-1,2,3 and 4}. The calculated spectrum, for each set of parameters, has sixteen eigenvalues which are associated
to physical meson-states, according to
their dominant flavor content and energy, as we shall discuss later on.
These energies and wave functions are used to calculate the widths of the states, in the manner described in the next sub-section.
The procedure is taken from Ref.\cite{Mottelson}
\subsection{The width of the states}\label{energy-width}
The Hamiltonian is written
\beqa\label{H}
H=H_0+V
\eeqa
where $H_0$ is the Hamiltonian of Eq.(\ref{H-RPA-flavor})
mapped onto the RPA basis 
\beqa
H_0=\sum_n E^{RPA}_n \gamma^\dag_n \gamma_n~,
\eeqa
with $n=1,\cdots 16$. The interaction term $(V)$, describes
the interactions not included in the RPA treatment \cite{Yepez2017}.
To calculate the width of a state $| a\rangle =\gamma^\dag_a | \tilde 0\rangle$ 
we assume that the basis can be separated
in a set of reference states $\{|a\rangle\}$
and a background $\{| \alpha \rangle\}$ with $N$ elements, such that \cite{Mottelson}
\beqa\label{elements}
H_0 | a \rangle &=& E_{a} | a \rangle\nonumber\\
H_0 | \alpha \rangle &=& E_{\alpha} | \alpha \rangle\nonumber\\
\langle a | V | a \rangle &=& 0\nonumber\\
\langle \alpha_j | V | \alpha_{j'} \rangle &=& V_{ \alpha_{j}, \alpha_{{j'} }}=0
~~~~\forall ~j,j' \nonumber\\
\langle a | V | \alpha_{j} \rangle &=& V_{a,\alpha_j}=V_{\alpha_j,a}
=\mbox{real} ~.
\eeqa
leading to the Hamiltonian-matrix
\beqa\label{matrix}
H=
\left( \begin{array}{cccccc}
E_{a} & V_{a,\alpha_1}     & V_{a,\alpha_2}& V_{a,\alpha_3} &\cdots&V_{a,\alpha_N} \\
V_{a,\alpha_1} & E_{\alpha_1} & 0                        &0                        & \cdots&0          \\
V_{a,\alpha_2} & 0                         &E_{\alpha_2} & 0                       &\cdots&0\\
 . &. &.&.&.\\
V_{a,\alpha_N} & 0                        &0                         &0                          &\cdots&E_{\alpha_N} \\
\end{array} \right) \, ,
\eeqa

Any eigenstate of the Hamitonian of Eq.(\ref{H}) can be written as
\beqa\label{eigenstate}
| E \rangle = c_{a}(E) | a\rangle 
+\sum_{j} c_{\alpha_j}(E)  | \alpha_j\rangle ~.
\eeqa

So that
\beqa
H | E \rangle &=& c_{a}(E) \left(H_0+V\right) | a\rangle 
+\sum_{j} c_{\alpha_j}(E) \left(H_0+V\right) | \alpha_j\rangle \nn\\
&=& E | E \rangle\nn\\
\eeqa
where
\beqa\label{ecs-sys}
\langle a | H | E \rangle=E  c_{a}(E) &=& 
 c_{a}(E) E_{a} + \sum_{j}  c_{\alpha_j}(E) V_{a,\alpha_j}  \nonumber\\
\langle \alpha_j | H | E \rangle=E  c_{\alpha_j}(E) &=& 
 c_{a}(E) V_{a,\alpha_j} + c_{\alpha_j}(E) E_{\alpha_j}  \nonumber\\
\eeqa
The above equations and the normalization condition $\langle E \mid E\rangle=1$ lead to the amplitudes
\beqa\label{amps}
c_{\alpha_j}(E)  &=&- c_{a}(E) \frac{ V_{a,\alpha_j} } { (E_{\alpha_j}-E)} \nn\\
\left( c_{a}(E)  \right)^2 
&=& 
\left( 1+ \sum_{j} \frac{ \left( V_{ a,\alpha_j } \right)^2}{ ( E_{\alpha_j }-E )^2 }\right)^{-1}~.
\eeqa
Then, the mean value of the energy, $\bar E$, and the width, $\Gamma$,
of the state with $E \approx E_a$ are given by the expressions \cite{Bes}
\beqa\label{mean}
\bar E &=& E_a \left(c_a(E)\right)^2 +\sum_j  E_{\alpha_j}\left(c_{\alpha_j}(E)\right)^2\nn\\
\Gamma&=&2\sigma
=2 \left( (E_a - \bar E)^2 \left(c_a(E)\right)^2 
+\sum_j  (E_{\alpha_j}-\bar E)^2\left(c_{\alpha_j}(E)\right)^2
\right)^\frac{1}{2}~.
\eeqa
%%%%%%%%%%%%%%%%%%%%%%%%%%%%%%%%%%%%%%%%%%%%%%%%%%%%%%%%%%%%
%%%%%%%%%%%%%%%%%%%%%%%%%%%%%%%%%%%%%%%%%%%%%%%%%%%%%%%%%%%%
%%%%%%%%%%%%%%%%%%%%%%%%%%%%%%%%%%%%%%%%%%%%%%%%%%%%%%%%%%%%
\section{Numerical Analysis of the solutions: energy and widths of  meson-like states}
\label{Sec3}
\begin{figure}[H]
\centering
\includegraphics[width=0.6\textwidth]{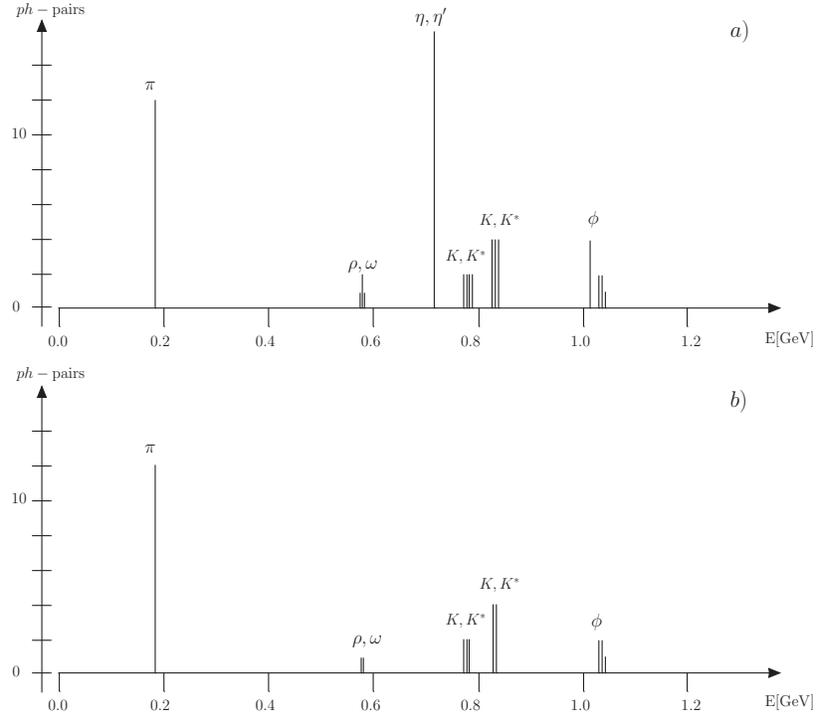}
\caption{Set-1: Structure of the RPA-eigenvalues, in terms of the number of particle(quark)-hole(antiquark) pairs. 
The upper inset a) shows the complete RPA spectrum. In the lower inset b) the composition of the background-states is shown.}\label{collec-1}
\end{figure}

\begin{figure}[H]
\centering
\includegraphics[width=0.6\textwidth]{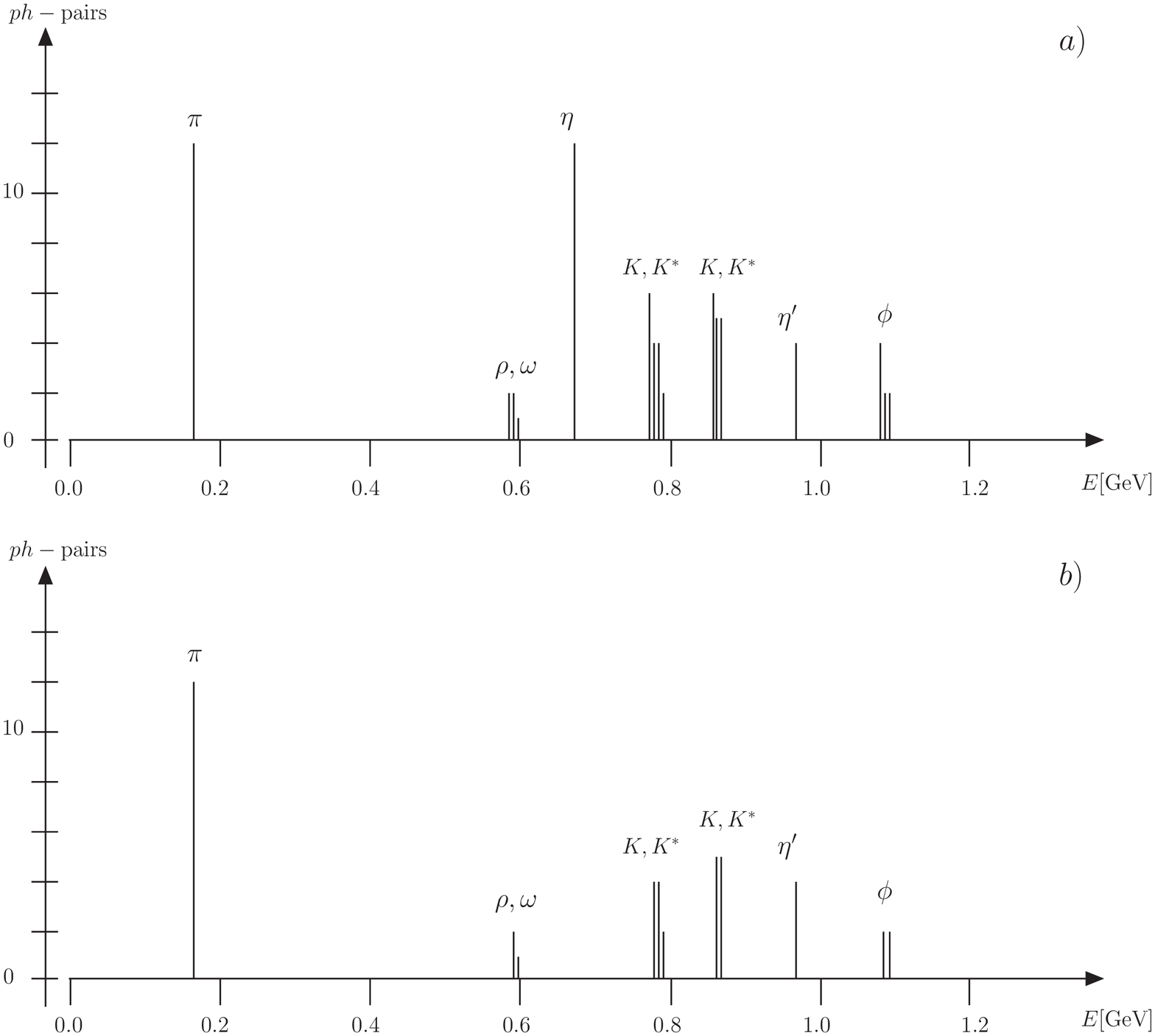}
\caption{Same as Figure 1, for the set 2 of parameters.}\label{collec-2}
\end{figure}

\begin{figure}[H]
%\begin{figure}[!h]
\centering
\includegraphics[width=0.6\textwidth]{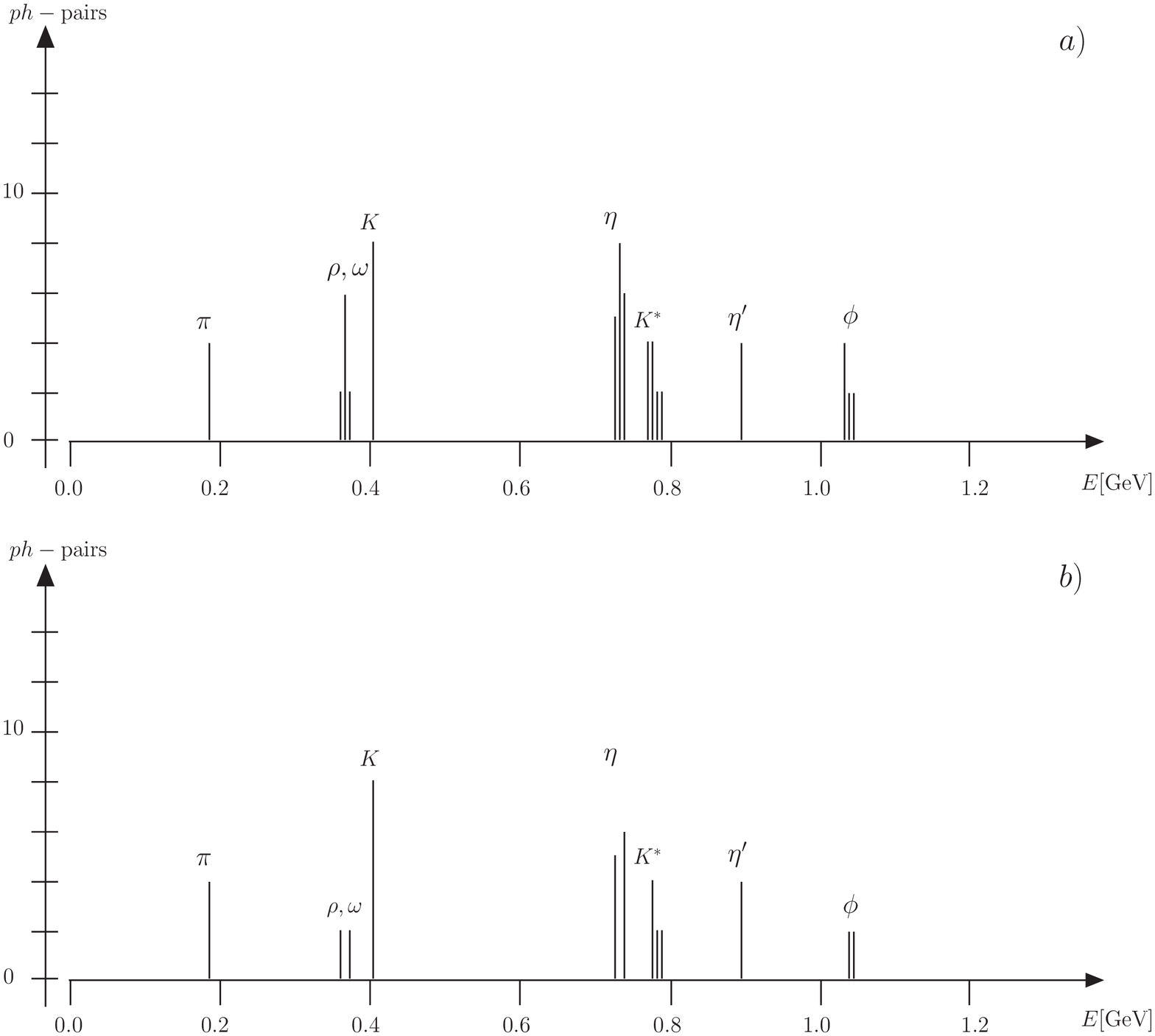}
\caption{Same as Figure 1, for the set 3 of parameters}\label{collec-3}
\end{figure}

\begin{figure}[H]
%\begin{figure}[!h]
\centering
\includegraphics[width=0.6\textwidth]{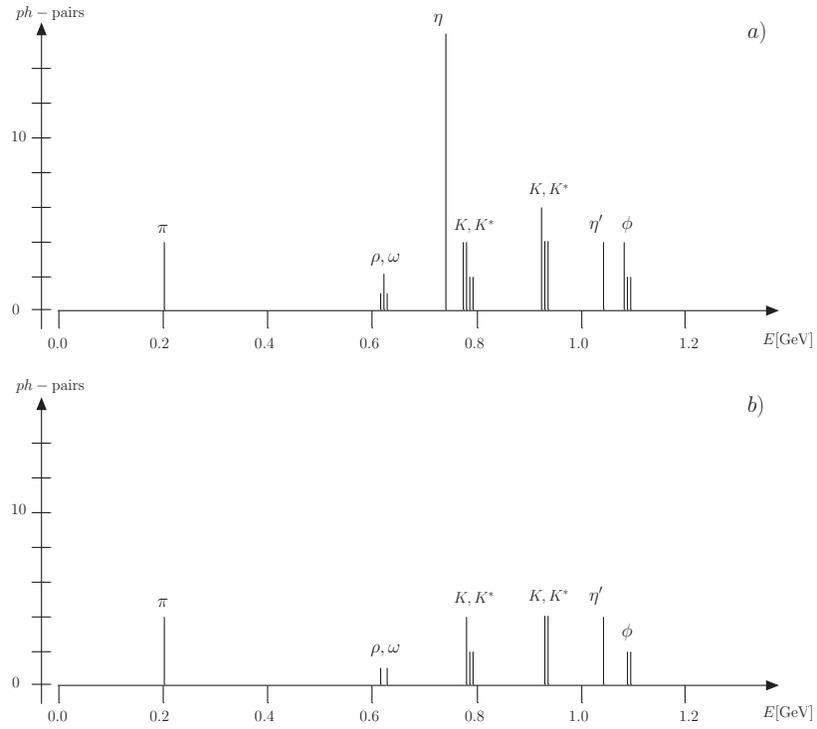}
\caption{Same as Figure 1, for the set 4 of parameters.}\label{collec-4}
\end{figure}

The low-energy scalar-meson-states have large widths \cite{PDG2016}. This is the case of the
state tentatively identified as $f_0(500)$ (or $\sigma$) with a width
of about $400-700$ MeV. The existence and the structure of this scalar-meson state have been rather controversial  
since it could be interpreted as a four-quark state or as a two-meson molecule \cite{Albaladejo2012}. 
In \cite{Amor2017} we were able to identify scalar-mesons as solutions of a non-perturbative 
approach based on the use of many-body methods. However,
scalar meson-states are beyond the reach of the minimal SO(4)-model developed
in \cite{Yepez2016,Yepez2017} since angular or radial
excitations are needed to get quark-antiquarks meson-like states of positive parity. 
For pseudoscalar mesons up to 1 GeV the data 
indicate that they have narrower widths while broader widths are reported for vector mesons. 
That is the case of the $\rho$-meson. In Table \ref{tab1} we list the values taken from \cite{PDG2016}.
\begin{center}
\begin{table}[h!]
\centering
\begin{tabular}{c| cc c cc c cc c cc c cc c cc c}
\hline\hline
Width \textbackslash State && $\eta$&& $\rho$&&$\omega$ &&$K^*$&& $\eta'$ && $\phi$ \\ [0.5ex] 
\hline
$\Gamma$  &&1.3  && 147.8 &&8.5  &&50.8 &&0.2&&4.3 \\ [0.5ex] 
\hline\hline
\end{tabular}
\caption{Observed values of the widths $\Gamma$ of pseudoscalar and 
vector mesons. The values are given in units of [MeV] and they have been taken from Ref.\cite{PDG2016}.}
\label{tab1} 
\vspace{0.2cm}
\end{table}
\end{center}
The width of the $\eta'(957)$ state is about $0.02\%$ of the mass of the state. Because of this rather small value 
it will be considered as an isolated state. For the rest of the states their
widths vary between $\approx 1\% \to 20\%$ of
their masses and they will be calculated using the formalism presented in the previous section. 
In the following we shall perform a case by case analysis of the results obtained with each set of parameters of the Hamiltonian of Eq.(\ref{H-RPA-flavor}). They are  
listed in Table \ref{Sets}. The meaning of this parametrization, their values and the effects of it upon the meson spectrum have been discussed in detail in Refs.
\cite{Yepez2016,Yepez2017}.
\begin{center}
\begin{table}[h!]
\centering
\begin{tabular}{c | cc c cc c cc c cc c c c c c  cc c cc  c  cc  c  cc c cc c}
\hline\hline
%ind&&
set &&$a_1$ && $a_2$ &&$a_3$&& $a_5$&&$a_6$ &&$a_7$ &&$b$ 
\\ [0.5ex] \hline
1&&100 && 50 && 200 && -300 && 100 && -150 && 45.00
\\[0.5ex] 
2&&100 && 100 && 200 && 0 && 50 && -50 && 58.12
\\[0.5ex] 
3&&100 && -100 && 200 && 0 && 100 && -150 && 54.37
\\[0.5ex] 
4&&100 && 150 && 200 && 0 && 0 && 0 && 54.37
\\[0.5ex] 
\hline
\hline  
\end{tabular}
\caption{Parameters of the Hamiltonian  of Eq.(\ref{H-RPA-flavor}). The values are given in units of $[\mbox{MeV}]$. }\label{Sets} \vspace{0.2cm}
\end{table}
\end{center}
We have solved the RPA-eigenvalue problem and 
classified the eigenvectors by inspecting their flavor content in order to established a correspondence between the
RPA spectrum and physical states. The results of such a procedure are given in Table \ref{rpa-eigen}.
\begin{center}
\begin{table}[h!]
\centering
\begin{tabular}{c| cc c cc c cc c cc c cc c}
\hline\hline
State && set 1 && set 2 && set 3 && set 4 \\ [0.5ex] 
\hline
$\pi$  &&184.81 (1) && 164.49 (1) && 187.27 (1)&& 201.37 (1) \\ [0.5ex] 
\hline
$\rho$ $\omega$  && 579.10 (3)&& 590.31 (3) && 364.98 (3)&&622.56 (3) \\ [0.5ex]
\hline
$\eta$  &&716.44 (1)  && 670.06 (1) &&735.12 (3) && 741.59 (1) \\ [0.5ex]
\hline
$\eta^{\prime}$  && && 965.95 (1)&&895.17 (1) &&1042.88 (1) \\ [0.5ex]
\hline
$K$  $K^{*}$ (low)  &&780.00 (4)  && 780.00 (4) && 402.11 (1) &&780.00 (4)\\ [0.5ex]
\hline
$K$  $K^{*}$    (high)  &&827.55 (3)  && 863.27 (3) &&780.00 (4)  &&930.00 (3) \\ [0.5ex]
\hline
\multirow{2}{*}{$\phi$}  &&1011.00 (1) && \multirow{2}{*}{1086.41 (3)}&&\multirow{2}{*}{ 1039.89 (3)}  &&\multirow{2}{*}{1087.43 (3)} \\ [0.5ex]
 &&1033 (3)  &&  && &&  \\ [0.5ex]
\hline\hline
\end{tabular}
\caption{RPA energies, in units of MeV, for the eigenvectors associated to physical states \cite{PDG2016} accordingly to the structure of their wave functions. 
The values in parenthesis indicate the degeneracy of each state. The sum of the degeneracies of each set equals the number of eigenvalues of the RPA basis.}\label{rpa-eigen} 
\vspace{0.2cm}
\end{table}
\end{center}
%%%%%%%%%%%%%%%%%%%%%%%%%}
Due to the SO(4) symmetry of the Hamiltonian the $\rho$- and $\omega$-like states appear as a mixture,
as well as the kaon ($K,K^*$)-like states \cite{Yepez2016}. The
breaking of this degeneracy is beyond the SO(4) 
scheme since it requires the inclusion of radial and orbital excitations. 
% This is also the case of the $\eta$ and $\eta^{\prime}$ mesons. 
As seen from the results listed in Table \ref{rpa-eigen} for the set 1 
there was only one possible state, at 716.44 MeV, that
resembles the flavor structure of the $\eta$ and $\eta'$ states. 
We have assumed that these states are degenerate. The $\phi$-like states obtained for
set-1, show a small energy difference of about 22 MeV. For the calculations we will
consider that the state at 1011 MeV represents the physical meson $\phi(1020)$ state, and that the
other $\phi$-like-states belong to the background. The set-3 gives in the kaon-like sector of the spectrum 
one state at low energy (402.11 MeV)  as
compared with the results obtained with the other sets of
parameters. This state resembles more likely the pseudoscalar kaon
and it will be not considered for the width analysis.
For the rest of the kaon-like states of {\it Set-1,2,3,4}, the
formalism presented in the previous section will be implemented in
order to determine their widths. Concerning the matrix elements of the interaction $V$, that is the elements $V_{a,\alpha}$ of Eq.(\ref{matrix}), they are taken as constants for each of the sets of parameters of Table \ref{Sets}. The magnitude of these constants depends upon the states, and their values are listed in Table \ref{widths2}
With these elements we have calculated the width of the states.
The results are shown in the Table \ref{compiled}.
%%%%%%%%%%%%%%%%%%%%%%%%%%%%%%%%%%%%
%%%%%%%%%%%%%%%%%%%%%%%%%%%%%%%%%%%%%
\begin{center}
\begin{table}[h!] 
\centering
\begin{tabular}{c| cc c cc c cc c cc c cc c}
\hline\hline
State && set 1 && set 2 && set 3 && set 4 \\ [0.5ex] 
\hline
$\rho$ $\omega$  && 147.84&&147.46&&147.14&&147.23\\ [0.5ex]
\hline
$\eta$  &&1,32&&1.32&&1.30&&1.32\\ [0.5ex]
\hline
$K$  $K^{*}$ (low)  &&50.82&&50.90&&  && 50.63\\ [0.5ex]
\hline
$K$  $K^{*}$    (high)  && 50.74&&50.81&&50.99&&50.99\\ [0.5ex]
\hline
$\phi$  &&4.30&&4.32&&4.29&&4.32 \\ [0.5ex]
\hline\hline
\end{tabular}
\caption{Calculated width $\Gamma$ (Eq.\ref{mean}) of the states, in units of MeV. The values have been obtained as described in the text. The result quoted for Set 3, sector $K,K^*$(high)  corresponds to the vector state since Set 3 distinguish the $K$ state from the  $K^*$ state.}\label{compiled} 
\vspace{0.2cm}
\end{table}
\end{center}
%%%%%%%%%%%%%%%%%%%%%%%%%%%%%%%%%%%%%%%%%%%%%%%%%%%
%%%%%%%%%%%%%%%%%%%%%%%%%%%%%%%%%%%%%%%%%%%%%%%%%%%
Following with the use of the formalism of the previous section we have calculated the energy centroids and the average
interaction for each of the states. The average value of the interaction which produces the broadening of the states
is shown in Table \ref{widths2}.
\begin{center}
\begin{table}[h!]
% \begin{tabular}{lSSSSSSSS}
 \begin{tabular}{c| cc c cc c cc c cc c cc c  cccc}
\hline\hline
    \toprule
    \multirow{2}{*}{State} &
      \multicolumn{4}{c}{set 1} &
      \multicolumn{4}{c}{set 2} &
      \multicolumn{4}{c}{set 3} &
      \multicolumn{4}{c}{set 4} \\
      &&$\bar E$ && $V$ &&$\bar E$ && $V$ &&$\bar E$ && $V$ &&$\bar E$ && $V$ \\
      \midrule
\hline
$\rho$ $\omega$  && 598.91 && 45.15&& 610.43&&45.5&& 367.87 && 34.5 && 645.69 && 50.00\\ [0.5ex]
\hline
$\eta$  &&716.44&&0.20&&670.06&&0.20&&735.12&&0.29&&741.59&&0.20\\ [0.5ex]
\hline
$K$  $K^{*}$ (low)&& 787.03&&14.05&&783.11&&12.80&& &&  &&781.36&&12.60\\ [0.5ex]
\hline
$K$  $K^{*}$    (high)  && 825.80&&12.35&&862.75&&12.10&&779.61 &&12.30 &&930.36&&12.10\\ [0.5ex]
\hline
$\phi$  &&1011.04 && 0.65&&1086.39 && 1.02&&1039.88 && 0.96&&1087.40 && 1.02\\ [0.5ex]
    \bottomrule
\hline\hline
  \end{tabular}
\caption{Energy centroids $\bar E$ and parametrized interaction energy $V$, in units of MeV, for each of the 
sets of parameters considered in the calculations.} \vspace{0.2cm}
\label{widths2}
\end{table}
\end{center}
As said before the calculation of the width depends upon the choice of physical and background states and the nature
of each state is being determined by the composition of its wave function in terms of quark-antiquark pairs.
To give an idea about the structure of the RPA eigenvalues
in Figures \ref{collec-1}-\ref{collec-4} we show the collectivity of the states obtained with the different sets of parameters.
The corresponding amplitudes are represented by the number of pairs which contributed to each meson-like
state.  
By comparing the data of Table \ref{tab1} with the calculated values listed in Table \ref{compiled} we see that the agreement is
quite remarkable considering the rather simple structure of the Hamiltonian, which is the SO(4) version of the QCD Hamiltonian in the Coulomb Gauge. In all cases the order of magnitude is correct and the sensitivity of the calculations respect to the structure of the light-meson states is very strong since it correlates one by one with the physical states. The calculated values are indeed quite good, particularly in view of the huge variation of the data, which for the $\eta$-meson assigns a width of the order of 1.3 MeV and in the other extreme assigns much larger value (147.8 MeV) to the $\rho$-meson. 

The features shown by the numerical results are indeed supported by the analytical 
solutions of the model, e.g: the ones obtained by using an average interaction proportional to the average energy spacing and degeneracies of the states. 
From the grouping of states around a given reference state, shown in Figures \ref{collec-1}-\ref{collec-4}. 
it is possible to extract an average interaction and
take the analytic limit of the model, e.g:  one state merged in the background. The results of such a calculation yield values of the interaction quite similar to these of Table \ref{widths2}.

\section{Summary}\label{Sec4}
In this work we have extended the study of our previous publications  concerning the 
treatment of the QCD Hamiltonian in the Coulomb Gauge. We have taken the dominant sector of it as to be represented by the generators
of the SO(4) symmetry and parametrized the structure of the Hamiltonian in terms of the Casimir operators of the group, in order to calculate the spectrum of light-meson states. The Hamiltonian was diagonalized by applying the RPA method, which yields eigenvalues 
whose eigenvectors could be associated to physical states after analysing their composition in terms of quark and antiquark pairs \cite{Yepez2016,Yepez2017}.
In order to test these wave functions we have calculated the energy-width of each state by letting them to interact with a background
of less-collective or non-collective states. We have found that the calculated values do agree with data, for the four sets of parameters considered in the calculations. Thought the spectrum depends smoothly upon the parameters of the Hamiltonian, the calculation of the width of the states is parameter-free once the spectrum of physical and background states is properly defined. From these results we conclude that the identification of the states which we have performed, by looking at their particle-hole content, is physically sound. Therefore 
the procedure may be applied to more involved situations, like the one of \cite{Amor2017}, where the RPA treatment of the QCD Hamiltonian in the Coulomb Gauge was not restricted to the SO(4) limit.

\section*{Acknowledgments}
One of the authors (T.Y-M) thanks the National Research Council of
Argentina (CONICET) for a post-doctoral scholarships. (O.C.) is a member 
of the scientific career of the CONICET. (P.O.H) acknowledges financial help from
DGAPA-PAPIIT (IN100315) and from CONACYT (Mexico, grant 251817). 
This work has been supported financially
by the CONICET (PIP-282) and by the ANPCYT of Argentina.

 \vfil
\end{document}